\documentclass[prl,twocolumn,showpacs,superscriptaddress,preprintnumbers,amsmath,amssymb]{revtex4-1}
\usepackage{graphicx}
\usepackage{dcolumn}
\usepackage{bm}
\usepackage[colorlinks=true,linkcolor=blue]{hyperref}
\usepackage{CJK,CJKnumb,CJKulem}
\begin{document}
\begin{CJK*}{GBK}{com}
\title{Local Rotational Jamming and Multi-Scale Hyperuniformities \\ in an Active Spinner System}
\author{Rui Liu}
\email{lr@iphy.ac.cn}
\affiliation{Beijing National Laboratory for Condensed Matter Physics and CAS Key Laboratory of Soft Matter Physics, Institute of Physics, Chinese Academy of Sciences, Beijing 100190, China}
\author{Jianxiao Gong}
\affiliation{National Center for Nanoscience and Technology, Beijing 100190, China}
\author{Mingcheng Yang}
\author{Ke Chen}
\affiliation{Beijing National Laboratory for Condensed Matter Physics and CAS Key Laboratory of Soft Matter Physics, Institute of Physics, Chinese Academy of Sciences, Beijing 100190, China}
\affiliation{School of Physical Sciences, University of Chinese Academy of Sciences, Beijing 100049, China}
\date{\today}
\begin{abstract}
An active system consisting of many self-spinning dimers is simulated, and a distinct local rotational jamming transition is observed as the density increases. In the low density regime, the system stays in an absorbing state, in which each dimer rotates independently subject to the applied torque. While in the high density regime, a fraction of the dimers become rotationally jammed into local clusters, and the system exhibits spinodal-decomposition like two-phase morphologies. For high enough densities, the system becomes completely jammed in both rotational and translational degrees of freedom. Such a simple system is found to exhibit rich and multiscale disordered hyperuniformities among the above phases: the absorbing state shows a critical hyperuniformity of the strongest class and subcritically preserves the vanishing density-fluctuation scaling up to some length scale; the locally-jammed state shows a two-phase hyperuniformity conversely beyond some length scale with respect to the phase cluster sizes; the totally jammed state appears to be a monomer crystal, but intrinsically loses large-scale hyperuniformity. These results are inspiring for designing novel phase-separation and disordered hyperuniform systems through dynamical organization.
\end{abstract}
% \pacs{45.70.Qj}
\maketitle
\end{CJK*}

% \section{Introduction}
Disordered hyperuniform systems are exotic states of matter, which are isotropic as liquids or glasses but suppress long-wavelength density fluctuations as crystals \cite{Torquato2018,Torquato2003,Torquato2016a,Torquato2016b,Torquato2021a}. Amorphous materials with density hyperuniformity are naturally endowed with superior physical properties \cite{Florescu2009,Leseur2016,FroufePerez2017,Gkantzounis2017,Chen2018,Yu2021} and are thus of great importance in both science and technology. Synthesis of hyperuniform amorphous materials is a double challenge, as crystallization needs to be avoided while hyperuniformity should not be hampered. Single-component synthesis, which provides the great advantage of easy manipulation, hardly produces both disorder and hyperuniformity, especially in low dimensions.

Active matter may be promising for tackling such a difficulty. First, active systems may form structures due to dynamical organization which may avoid crystallization even for a monodisperse constituent. Monodisperse self-spinning rods have been observed to form some disordered structures though not discussed in detail \cite{Kirchhoff2005,vanZuiden2016}. Second, active matter utilizes energy locally, and thus may prevent the formation of meta-stable local structures which would destroy density hyperuniformity. Many active systems thus exhibit density hyperuniformity without relying on long-range interactions as required in equilibrium systems \cite{Torquato2018,Torquato2021}. Circle swimmer \cite{Lei2019a,Huang2021} and active spinner \cite{Lei2019b} systems have been shown to be density hyperuniform in their active fluidic regimes.  

Here by employing the previously studied model \cite{vanZuiden2016,Lei2019b}, we show that fast-spinning active dimers, not only undergo the reported jamming transition from an absorbing state to an intermediate squeezed fluid and finally to a completely-jammed state \cite{Kirchhoff2005,vanZuiden2016}, but also exhibit a distinct spinodal-decomposition like feature that jamming happens locally and homogenously. We identify that such an active spinner system in a critical absorbing state could also be both disordered and strongly hyperuniform, and the local jamming transition even leads to a two-phase hyperuniformity. 

The absorbing state subcritically preserves the hyperuniformity density-fluctuation suppression up to some length scale, which diverges at the critical point. The intermediate two-phase fluid shows a hyperuniformity scaling beyond some length scale with respect to the phase cluster sizes. The totally-jammed state appears to be a crystal for the constituent monomers of the dimers but an amorphous media for the dimer centers, which suppresses density fluctuations as a strongly hyperuniform case at small length scales but shows a non-hyperuniform characteristic at large scales. To our knowledge, such a rich exhibition of multiscale hyperuniformities in a single system is rare. 

%%%%%%%%%%%%%%%%%%%%%%%%%%%%%%%%%%%%%%%%%%%%%%%%%%%%%%%%%%%%%%%%%%%%
% \section{Simulation}
We simulate a two-dimensional (2D) active spinner system as those described in Ref.\cite{vanZuiden2016,Lei2019b}. Each spinner is a dimer consisting of two spherical monomers bonded with fixed length $\sigma=1$. Each monomer has a mass $m=1$ and each dimer has a moment of inertia $I=\frac{1}{2}m\sigma^2$. The monomers from different dimers interact with each other through a Weeks-Chandler-Andersen potential $U(r_{ij})=4\epsilon[(\frac{\sigma}{r_{ij}})^{6}-(\frac{\sigma}{r_{ij}})^{12}]+\epsilon$, when their separation $r_{ij}$ is smaller than a cutoff distance $r_c\equiv2^{1/6}\sigma$. With the total effect of all such pair interactions denoted by $U(t)$, the dynamics of any dimer $i$ is governed by 
\begin{gather}
2m\ddot{\bf r}_i=-2\gamma_t\dot{\bf r}_i-\nabla_i U(t)+\xi_i(t),\\
I\ddot{\theta}_i=\mathcal{T}-\gamma_s\dot{\theta}_i-\partial_{\theta_i}U(t)+\eta_i(t).
\end{gather}

By taking $\epsilon=1$ energy unit, and $\tau=\sqrt{m\sigma^2/\epsilon}$ to be the time unit, we set the driving torque $\mathcal{T}=1\epsilon$, the translational viscosity $\gamma_t=10 m/\tau$, and the rotational viscosity $\gamma_s=0.01\epsilon\tau$ (the corresponding relaxation time ratio $\tau_s/\tau_t = 1000$, which is crucial for the jamming transition to occur locally) in our simulation. The timestep of the simulation is set to $0.001\tau$, which is one percent of the translational timescale $\tau_t$. $\xi(t)$ and $\eta(t)$ are respectively the stochastic force and torque due to thermal fluctuations, which are set to 0 in this paper, as we only presents the zero-temperature data here. However, the results remains qualitatively the same for a range of low temperatures, which is very similar to that of the circle swimmers as discussed in Ref.\cite{Lei2019a}. For $N$ dimers in a square box with length $L$, the number density is simply evaluated as $\rho=N\sigma^d/L^d$ (with the dimension $d=2$). Periodic boundary conditions are applied for all box boundaries.

\begin{figure}
\includegraphics[width=0.95\linewidth]{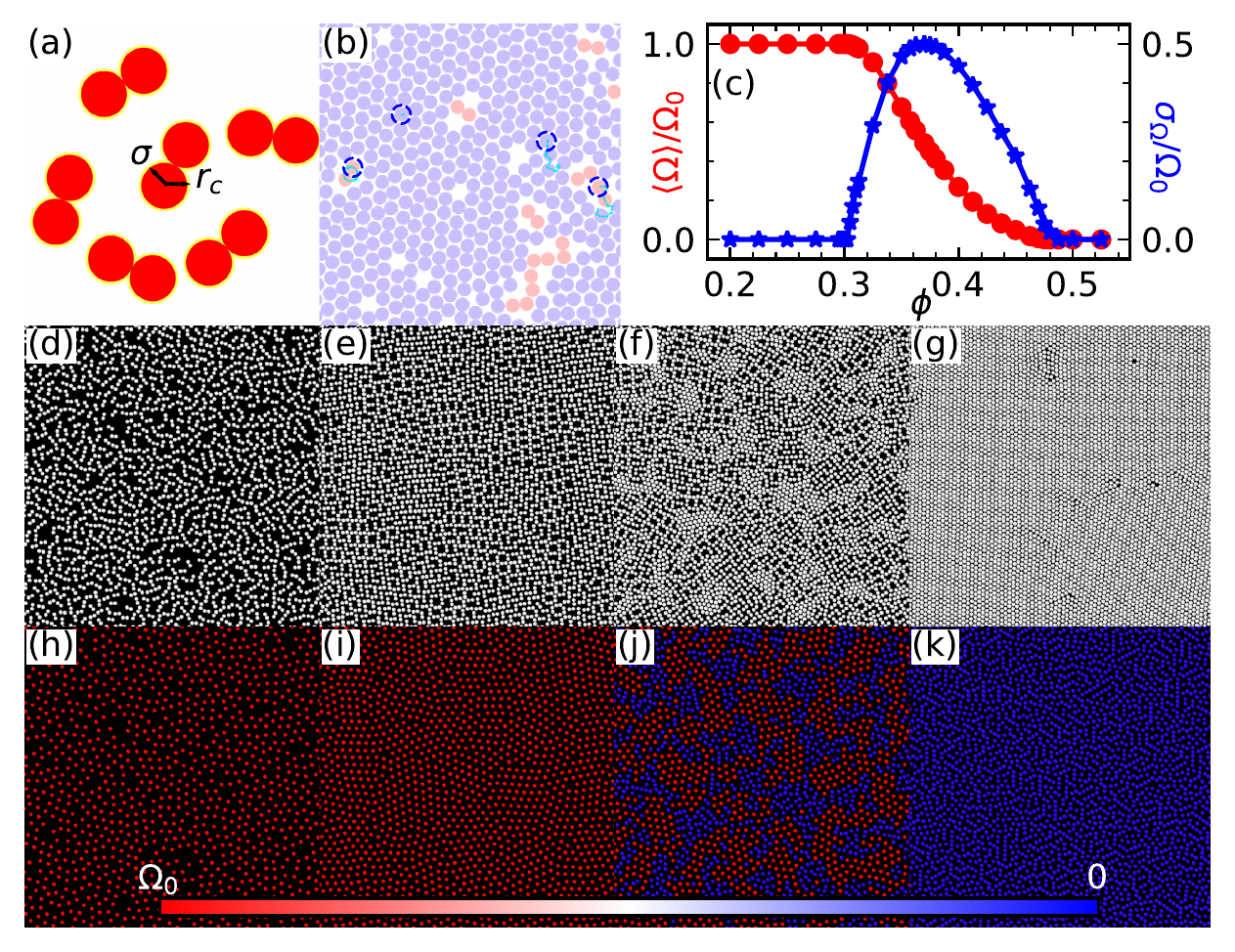}
\caption{\label{Fig.1}(a) Simulated dimers with core size parameter $\sigma$ and interaction distance cutoff $r_c=2^{1/6}\sigma$; (b) Four types of particle dynamics: monomers circled by blue dashes show different trajectories (cyan), long trajectories indicate diffusion while point/circle trajectories indicate jamming/absorbing, dimers in shaded red rotate freely and those in shaded blue barely rotate; (c) Average angular velocity $\langle\Omega\rangle$ and its standard deviation $\sigma_\Omega$, as functions of $\phi$; (d-g) Steady state snapshots for systems at $\phi=0.2,0.3,0.4,0.5$; (h-k) Dimer centers extracted from (d-g), with the spin velocity $\Omega$ indicated by colors.}
\end{figure}

The dimers in our simulation are much of a dumbbell shape [see Fig.\ref{Fig.1}(a)] as that in Ref. \cite{Lei2019b}, rather than the short-rod like shape in Ref. \cite{vanZuiden2016}. Thus, the long-range crystalline order observed in \cite{vanZuiden2016} is avoided in our case. On the other hand, we simulate a system much denser than that of \cite{Lei2019b}, which triggers a jamming transition and is thus conversely comparable to the results of \cite{vanZuiden2016} or the long-rod version in \cite{Kirchhoff2005}. However, hyperuniformity properties of such spinner systems have not been explored in \cite{Kirchhoff2005,vanZuiden2016}, and have been investigated only in the active fluidic regime in \cite{Lei2019b}. Here, we reveal a strong hyperuniformity of the critical absorbing state of the spinner system at the onset of a local rotational jamming transition, and further show the development of a two-phase hyperuniformity beyond the critical point.

With the above settings, the simulated system is found to exhibit three states. First, an absorbing state, in which each dimer rotates independently subject to the applied torque, is observed for relatively low densities [Fig.\ref{Fig.1}(d)]. As the density increases, the system undergoes a jamming transition, which occurs locally at first. The main jamming feature currently observed is in the rotational degree of freedom that the involved dimers cease to spin freely and are compacted into local clusters. As the remaining dimers still keep spinning and are organized into relatively loose structures, a spinodal-decomposition like two-phase morphology develops [Fig.\ref{Fig.1}(f)]. For a very dense system, no dimer can diffuse and rotate any more, and the system gets totally jammed in all degrees of freedom. This final jammed state appears to be an imperfect crystal for the monomers, but actually creates an amorphous configuration for the dimer centers [Fig.\ref{Fig.1}(g)]. 

Generally, four typical types of particle dynamics can be observed during the whole procedure, as indicated by blue circles in Fig.\ref{Fig.1}(b): (1) absorbing but freely-spinning; (2) completely jammed; (3) rotationally-jammed but diffusing; (4) spinning and diffusing. The previous two are respectively the only type of particle dynamics in the absorbing state and that in the final jammed state, while all the dynamics can be observed in the intermediate locally-jammed state. The local jamming actually leads to a two-phase fluid, and the diffusive dynamics can be observed either near the phase boundaries or in clusters moving as a whole. 

%%%%%%%%%%%%%%%%%%%%%%%%%%%%%%%%%%%%%%%%%%%%%%%%%%%%%%%%%%%%%%%%%%%%%%%%%%
% \section{Local Rotational Jamming Transition}
The local rotational jamming transition corresponds to a crossover in the averaged angular velocity of all the spinners $\bar{\Omega}=\langle\Omega\rangle/\Omega_0$ (where $\Omega\equiv\dot{\theta}$, and the measure is normalized by the freely-spinning velocity $\Omega_0=\mathcal{T}/\gamma_s$), as shown in Fig.\ref{Fig.1}(c). In this specific system, the crossover ranges from a lower transition point at $\phi\approx 0.31 $ to a higher one at $\phi\approx 0.48$.

Critically at the lower transition point $\phi=0.31$, cross structures (or T-structures called in Ref.\cite{Kirchhoff2005}) of freely-spinning dimers are observed Fig.\ref{Fig.1}(e). Due to spin phase differences, the dimers may not interfere each other, even when their centers get closer than the dynamical sweeping range $(\sigma+r_c)/2$. Any initial state of the system, with arbitrary overlaps between the dimers' sweeping ranges, would dynamically reorganize and spontaneously generate randomness in the spin phase differences. Such a mechanism automatically avoids crystallization and produces amorphous structures. Around the higher transition point $\phi=0.48$, a complete jamming trivially leads to a crystal-like structure at the monomer scale, with unavoidable dislocations generated by the self-imposed stresses and the odd elasticity \cite{Scheibner2020} of the jammed spinners. 

The local rotational jamming may start from a single dimer, which ceases spinning due to a little bit of overcrowding. As a result of the large relaxation time ratio $\tau_s/\tau_t= 1000$, the dimers tend to separate into distinct spinning and non-spinning regions rather than globally slow down, as the density increases. Since the squeezing occurs homogeneously, the system resembles a spinodal-decomposition process and presents a two-phase morphology, with high-density clusters of $\Omega\approx 0$ in the rotationally-jammed regions and freely-spinning cross-structures of $\Omega\approx\Omega_0$ for the rest. The above features ensure that the crossover corresponding to the transition never gets as sharp as that in Ref.\cite{Kirchhoff2005}.

As angular velocities of the dimers are spontaneously discretized into two values, statistics of $\Omega$ automatically measure the number faction of spinners in either phase for the two-phase morphology. The standard deviation of the angular velocity $\sigma_\Omega$ reaches its maximum value $\Omega_0/2$ at about $\phi=0.37$ [Fig.\ref{Fig.1}(c)], indicating an equal partition between the two phases, with half free spinners and half rotationally-jammed ones. In contrast with the crossover in $\bar{\Omega}$, one observes an abrupt change in $\sigma_\Omega$ at both the lower and higher transition points. The rapid deviation from 0 when $\phi$ increases from $0.31$ or decreases from $0.48$ shows that $\sigma_\Omega$ is a better indicator for the local jamming transition.

%%%%%%%%%%%%%%%%%%%%%%%%%%%%%%%%%%%%%%%%%%%%%%%%%%%%%%%%%%%%%%%%%%%%%%%%%%
% \section{Disordered Hyperuniformity}
\begin{figure}
\includegraphics[width=0.9\linewidth]{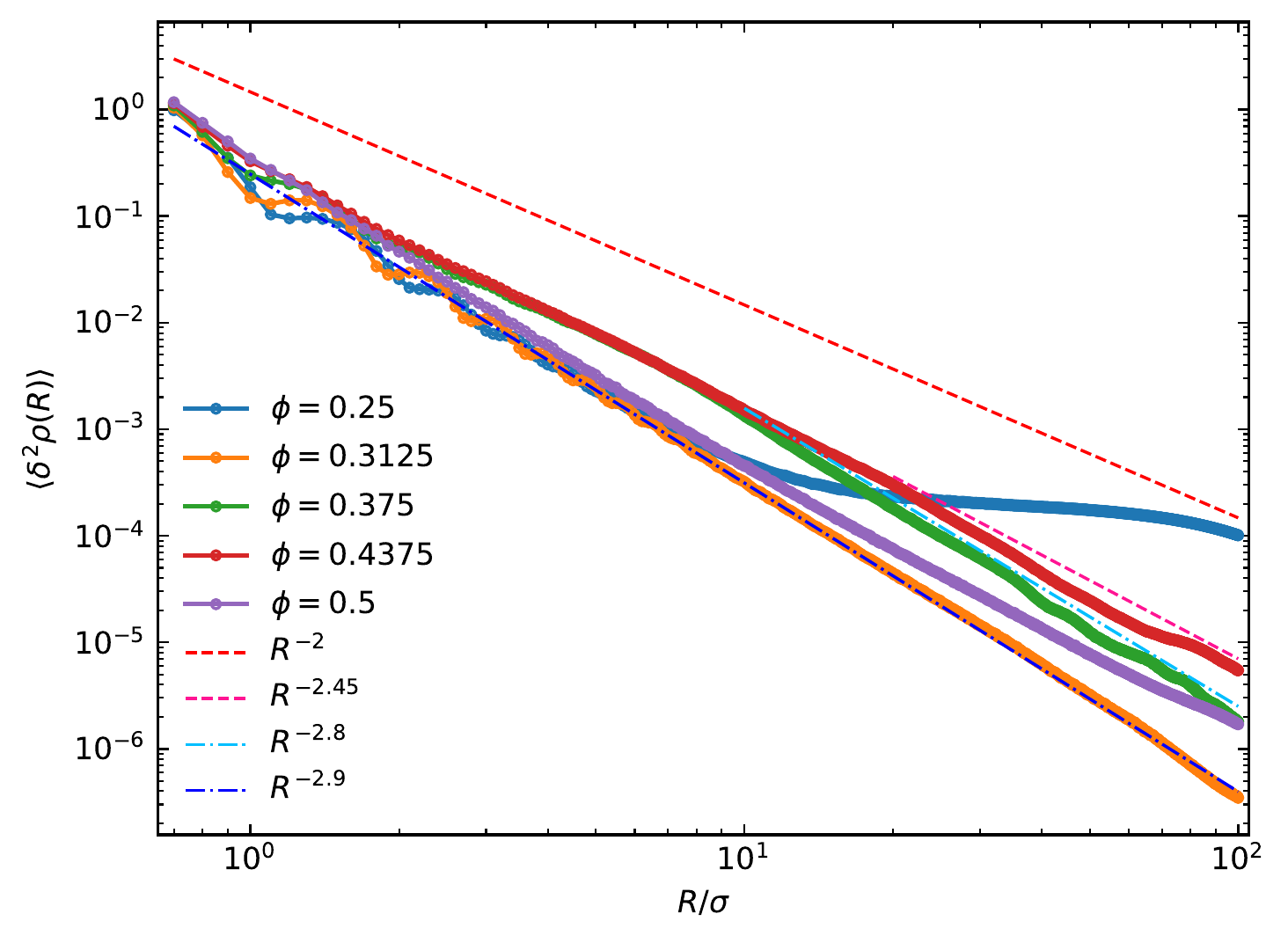}
\caption{\label{Fig.2}Density fluctuations $\langle\delta^2\rho\rangle$ as functions of window size $R$ for various $\phi$, measured at fixed system size $L=400\sigma$. Various lines show different asymptotic $R^{-\alpha}$ scalings.}
\end{figure}

The system produces disordered patterns for various densities, below or beyond the lower critical point $\phi\approx 0.31$. Even for the totally jammed states beyond the higher critical point $\phi\approx 0.48$, randomness in the orientation of the dimers causes strong disorder in their center-of-mass configurations. Steady-state configurations of the dimer centers for different states are respectively shown in Fig.\ref{Fig.1}(h-k), which are obviously disordered as point configurations. The rich disordered patterns exhibited by the system have different extents of uniformity, as can be seen from Fig.\ref{Fig.1}. Treated as point configurations, one can measure the density fluctuations $\langle\delta^2\rho(R)\rangle\equiv\langle\rho^2(R)\rangle-\langle\rho(R)\rangle^2$ for a randomly chosen spherical window with radius $R$ ($R<L/2$ to avoid finite size effect). The analysis can be done on either the monomers or the dimer centers, for either all spinners or only the free ones. Here in Fig.\ref{Fig.2}, we only present the analysis on dimer centers for all the spinners in the system. 

The absorbing state at a density below the lower critical point (e.g. $\phi=0.25$ in Fig.\ref{Fig.2}), shows a density fluctuation suppression with a vanishing scaling relation $\langle\delta^2\rho(R)\rangle\sim R^{-2.9}$ extending up to some length scale (about $10\sigma$ for $\phi=0.25$). Such a behavior commonly exists for absorbing states of many different systems \cite{Hexner2015,Zheng2021}. The length scale diverges as the density approaches the critical value $\phi\approx 0.31$, and the vanishing scaling relation extends to the system size scale (see $\phi=0.3125$ in Fig.\ref{Fig.2}). This result evidences that the critical absorbing state has hyperuniformity properties of almost the strongest class. 

The jammed state at density beyond the higher critical point (e.g. $\phi=0.50$ in Fig.\ref{Fig.2}) has small density fluctuations, which is comparable with the critical hyperuniform case even for large $R$. However, the large-$R$ asymptotic behavior $\langle\delta^2\rho(R)\rangle\sim R^{-2}$ indicates that the jammed configuration is intrinsically non-hyperuniform.

The locally jammed states corresponding to the crossover have diffusive dynamics, and we naively perform the same analysis for some instantaneous steady-state snapshots. The results for $\phi=0.375$ and $0.4375$ are shown in Fig.\ref{Fig.2}. For small $R$, the density fluctuation behavior mimics those of the absorbing and jammed states, since statistics are mainly done inside one phase and phase boundaries are rarely involved. Beyond some length scale (typically $10\sigma$), a scaling relation $\langle\delta^2\rho(R)\rangle\sim R^{-\alpha}$ with $\phi$-dependent $\alpha$ is observed. Obviously, $\alpha$ varies due to changes of the cluster sizes of either phase. For a finite simulation set, the exponent $\alpha$ measured could be as small as $2.4$ (still well beyond the hyperuniformity criterion $2$), or as large as $2.9$ (close to the strongest hyperuniformity).

\begin{figure}
\includegraphics[width=\linewidth]{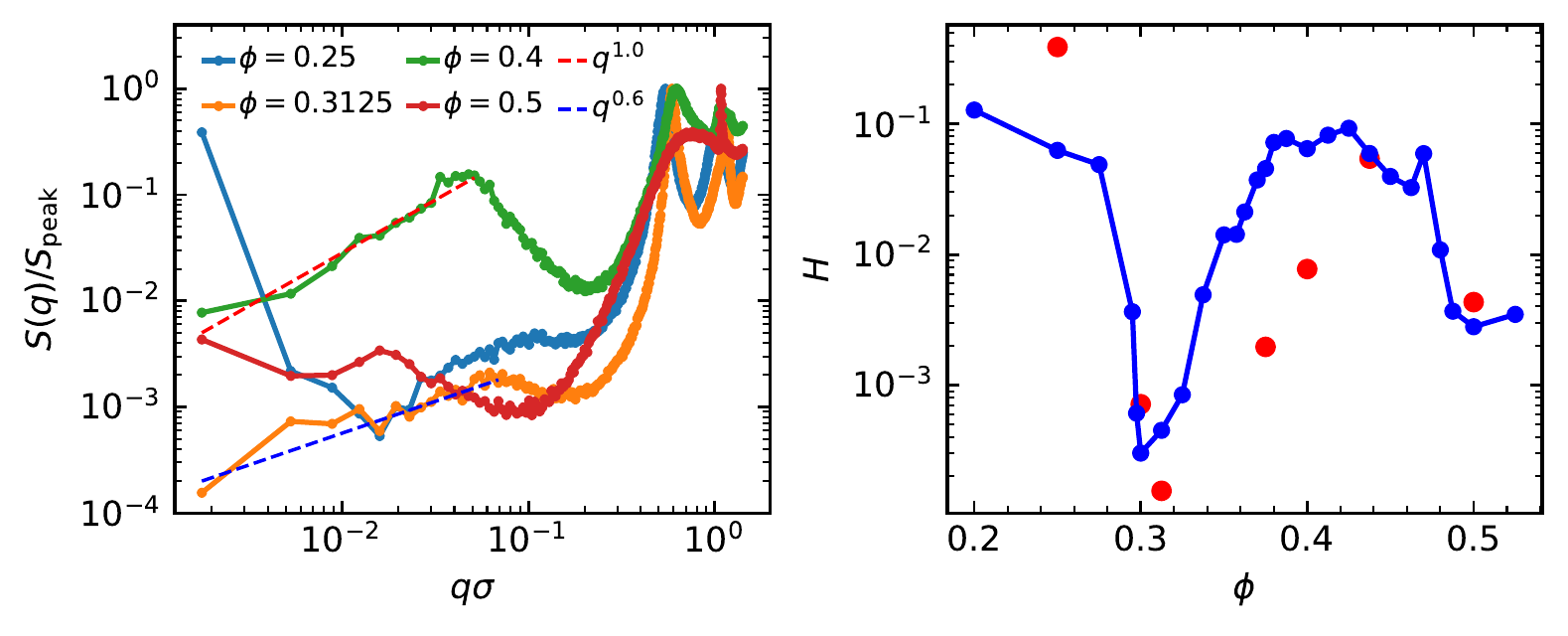}
\caption{\label{Fig.3}Structure factor $S(q)$ and hyperuniformity index $H(\phi)$: (a) $S(q)$ at $\phi=0.25,0.3125,0.4,0.5$ for system size $L=400\sigma$, the dashed lines show the asymptotic small-$q$ scalings; (b) The corresponding $H(\phi)$ to the system in (a) is shown in red, and much more data for a smaller system ($L=200\sigma$) is shown in blue for comparison.}
\end{figure}

To further characterize the disordered hyperuniformity of the system, we calculate the structure factor $S(q)$ for such point configurations of dimer centers, which are shown in Fig.\ref{Fig.3}(a). Typically for a hyperuniform system, $S(q)\rightarrow 0$ as $q\rightarrow 0 $, with asymptotic scaling $S(q)\sim q^\alpha \;(\alpha>0)$. The $S(q)$ for the subcritical absorbing state at $\phi=0.25$ or the supercritical jammed state at $\phi=0.5$ increases as $q\rightarrow 0$. These behaviors are consistent with the results of the density fluctuation analysis that the two states are not hyperuniform at large scales. For the critical absorbing state $\phi=0.3125$ and the locally-jammed state $\phi=0.4$, decays in $S(q)$ are found as $q\rightarrow 0$, respectively with scaling relations $S(q)\sim q^{0.6}$ and $\sim q^{1.0}$. 

A hyperuniformity index $H=S(q\rightarrow 0)/S_{\rm peak}$ (where $S_{\rm peak}$ is the peak value) can be defined with the calculated structure factors $S(q)$ \cite{Torquato2018}.  Empirically, one has $H\lesssim 10^{-3}$ for hyperuniform configurations \cite{Kim2019,Zheng2020}. We plot the corresponding $H$ as a function of $\phi$ in red for system of size $L=400\sigma$ in Fig.\ref{Fig.3}(b), and more data are shown in blue for a smaller system with $L=200\sigma$. In both cases, We observe a drastic decrease in $H$ to below $10^{-3}$ for the critical absorbing state, indicating a strong hyperuniformity. However, $H$ does not reflect the hyperuniform feature of the locally-jammed state, and is a bit misleading for becoming low in the supercritical jammed states which are intrinsically not hyperuniform.  

\begin{figure}
\includegraphics[width=\linewidth]{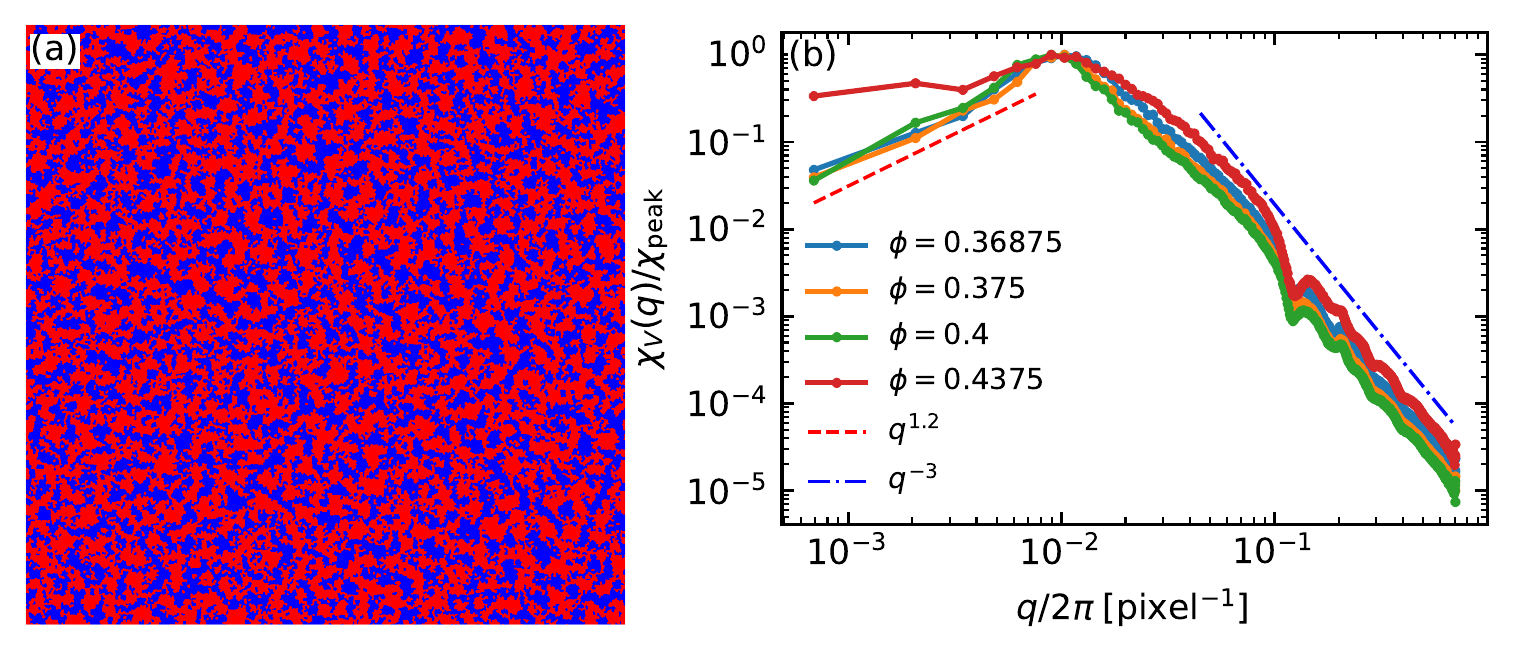}
\caption{\label{Fig.4}Two-phase hyperuniformity: (a) the two-phase field $\Omega({\bf r})$ for $\phi=0.375$; (b) the spectral density $\chi_V(q)$ for $\phi=0.36875,0.375,0.4,0.4375$, the dashed line at small $q$ shows a $q^{1.2}$ law approaching to zero, and the dash-dotted line shows a $q^{-3}$ scaling for large $q$.}
\end{figure}

%%%%%%%%%%%%%%%%%%%%%%%%%%%%%%%%%%%%%%%%%%%%%%%%%%%%%%%%%%%%%%%%%%%%%%%%%%
% \section{Two-phase hyperuniformity}
To correctly characterize the locally-jammed state, we need to treat it as a two-phase media, by thresholding some scalar field. The angular velocity $\Omega$ is a pseudo-scalar for a 2D system, one thus can construct a scalar field $\Omega({\bf r})$. We do this by simply apply a Voronoi tessellation to the point configurations of the locally-jammed states. The triangulated results are then mapped onto images with a fixed resolution of $2048\times 2048$ in pixels. As mentioned above, the values of $\Omega$ are automatically discretized into two values, $0$ and $\Omega_0$. Thus all vertexes in the tessellation diagram can be colored into either blue ($\Omega=0$) or red ($\Omega=\Omega_0$). One then can cut a triangle edge with both blue and red ending vertexes into halves, and connect the cut-points to make phase boundaries. The underlying pixel squares are then colored according to the major phase each contains. A typical two-phase field for $\phi=0.375$  is shown in Fig.\ref{Fig.4}(a).

For a hyperuniform two-phase media, its spectral density $\chi_V(q)$ approaches to zero as $q\rightarrow 0$, with $\chi_V(q)\sim q^\alpha$ ($\alpha>0$) \cite{Torquato2018,Torquato2016a}. For an image representation with square pixels, the spectral density can be measured as $\chi_V({\bf q})=m^2({\bf q})\phi^i S^i({\bf q})$ \cite{Chen2018,Kim2021}, where the shape factor $m({\bf q})={\rm sinc}(q_x/2){\rm sinc}(q_y/2)$, $\phi^i$ is the volume fraction for either phase denoted by the superscript $i$, and $S^i({\bf q})$ is the structure factor for the corresponding pixel centers.

We perform the analysis for the locally jammed states beyond the lower critical point, and the results are shown in Fig.\ref{Fig.4}(b). Obviously, such a two-phase state has a strong-hyperuniformity characteristics with $\alpha>1$ ($\alpha\approx1.2$). It is worth noting that all the normalized spectral density curves collapse except for some very large density as $\phi=0.4375$. This indicates that the relative phase cluster sizes, which generally change with $\phi$, are not critical for hyperuniformity. The system can stably develop hyperuniform morphologies for the two phases with either equal or unequal concentrations. The arbitrary shapes of the clusters also indicate that they possess little surficial energy. For $\phi=0.4375$, the small-$q$ scaling seems to have a much smaller $\alpha$, possibly due to some statistical reason. As we choose the active phase as the explicit one in all calculations, the number of the spinners involved in this case is quite small since most of them are jammed. Additionally, the system always has a $q^{-3}$ decay at the large $q$ side, which is similar to that of the numerically designed patterns through Fourier-space construction\cite{Chen2018}. 

\begin{figure}
\includegraphics[width=\linewidth]{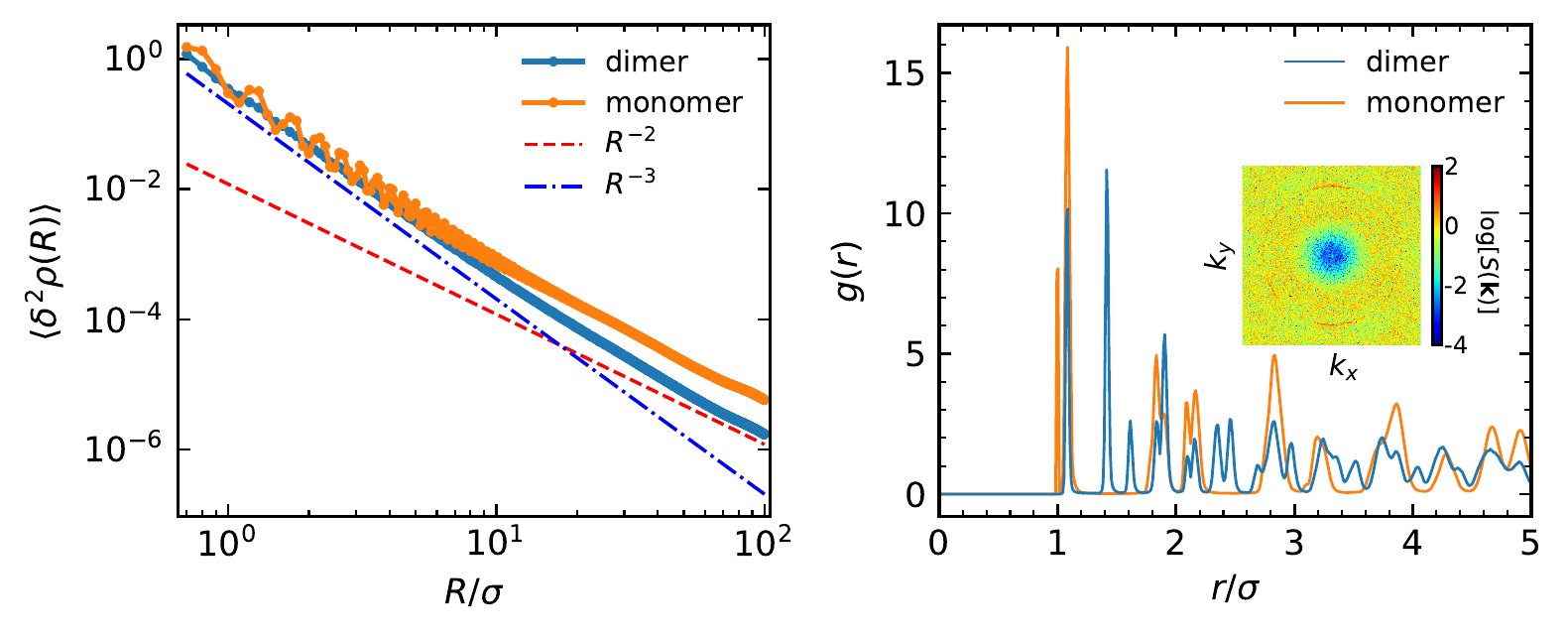}
\caption{\label{Fig.5}The totally jammed state at $\phi=0.5$: (a) density fluctuations $\langle\delta^2\rho(R)\rangle$ for the dimer centers (blue) and the monomers (orange), the dashed and dash-dotted lines respectively show the $R^{-2}$ and $R^{-3}$ scalings; (b) the radial distribution functions $g(r)$ for the dimer centers (blue) and the monomers (orange), the inset shows the structural factor $S({\bf k})$ for the dimer-center configuration.}
\end{figure}

%%%%%%%%%%%%%%%%%%%%%%%%%%%%%%%%%%%%%%%%%%%%%%%%%%%%%%%%%%%%%%%%%%%%%%%%%%
% \section{The Strange Jammed State}
As it is known that density hyperuniformity is related to incompressibility, the fact that the totally jammed state, which is nearly incompressible, is not hyperuniform, is strange, especially when the monomer configuration appears to be crystal-like. We perform the density-fluctuation analysis for the monomer configurations, which shows more-or-less the same behavior as that of the dimer centers, with $\sim R^{-3}$ at small $R$ and $\sim R^{-2}$ at large $R$ [See Fig.\ref{Fig.5}(a)]. We argue that such a behavior is due to some intrinsic constraints within the current system. The monomers form imperfect hexagons as we have $r_c\ne \sigma$, with the radial distribution function $g(r)$ peaking at both $r=\sigma$ and $r=\frac{\sigma+r_c}{2}$ [as shown in Fig.\ref{Fig.5}(b)]. Random orientations of the dimers leads to a random tiling of the imperfect hexagons, and we have an intrinsic fluctuation proportional to $\frac{r_c-\sigma}{2}$ in determining pair distances. Such a fluctuation is small but independent of the size $R$ of any window chosen for statistics. Thus, the fluctuation can not be averaged out and always causes a number-counting fluctuation proportional to the peripheral length of the window, i.e. $\sim R^{d-1}$. The fluctuation is neglectable for small $R$, but becomes dominant when the statistical fluctuation vanishes at large $R$. 

More interestingly, the \textit{disordered} dimer-center configuration, as the imperfect monomer crystal, also shows sharp peaks in $g(r)$ [Fig.\ref{Fig.5}(b)], and the corresponding structural factor $S({\bf k})$ [inset of Fig.\ref{Fig.5}(b)] still inherits a six-fold symmetry even though it has been somehow angularly blurred. The $S({\bf k})$, without an angular average, approaches zero at many but not all ${\bf k}$ points in some range like $\vert{\bf k}\vert<K$, which may be viewed as a variant originating from stealthy hyperuniformities \cite{Zhang2015,Torquato2015,Ma2017,DiStasio2018}. These strange features of the jammed state indicates that there is still plenty of room in-between ordered and disordered, hyperuniform or nonhyperuniform configurations, and thus are inspiring for designing new materials.

%%%%%%%%%%%%%%%%%%%%%%%%%%%%%%%%%%%%%%%%%%%%%%%%%%%%%%%%%%%%%%%%%%%%%%%%%%
% \section{Conclusions}
Conclusively, multi-scale disordered hyperuniformities are revealed in a simple active spinner system. The small requirements of the system on particle constitutions and pair interactions are encouraging for fabricating various hyperuniform amorphous structures through dynamical organization.

The critical hyperuniformity in an absorbing state of such spinner systems is uncovered for the first time, and is found to belong to almost the strongest class. A novel phenomenon also discovered is the subsequent two-phase hyperuniformity during the local jamming transition, which could also be of a strong class. Additionally, the final totally jammed state, which is intrinsically nonhyperuniform but with confined density fluctuations to a rather low level, may provide ideas for designing interesting nonhyperuniform or even antihyperuniform materials \cite{Torquato2021b,Oguz2019}.

Though only zero-temperature results are present here, we have done some finite temperature simulations. For low temperatures, all the results remain qualitatively the same, as discussed in Ref.\cite{Lei2019a}. When the temperature becomes high enough, the system goes into the active fluidic regime and a hyperuniform fluid as that in Ref.\cite{Lei2019b} is recovered.

Active spinner systems are known to break time-reversal symmetry and show topological behaviors \cite{Dasbiswas2018,Banerjee2017,Yang2021a,Yang2021b}. We have observed multiple vortices induced by caged diffusion of the free spinners in the two-phase regime, which obviously correspond to spontaneous topological edge flows along spontaneously generated phase boundaries. Furthermore, dislocations are found to inevitably exist due to ``odd" stresses of the active spinners in the totally jammed states. 

% \section{Acknowledgement}

\end{document}